\documentclass[jcp,twocolumn,superscriptaddress,floatfix]{revtex4}
\usepackage{graphicx}
\usepackage{amsmath}
\usepackage{amssymb}
\begin{document}
\title{Effective slippage on superhydrophobic trapezoidal grooves}

\author{Jiajia Zhou}
\affiliation{Institut f\"ur Physik, Johannes Gutenberg-Universit\"at
  Mainz, D55099 Mainz, Germany}
\author{Evgeny S. Asmolov}
\affiliation{A.N. Frumkin Institute of Physical Chemistry and
  Electrochemistry, Russian Academy of Science, 31 Leninsky Prospect,
  119071 Moscow, Russia}
\affiliation{Central Aero-Hydrodynamic Institute, 140180 Zhukovsky,
  Moscow region, Russia}
\affiliation{Institute of Mechanics, M.V. Lomonosov Moscow State
  University, 119991 Moscow, Russia}
  \author{Friederike Schmid}
\affiliation{Institut f\"ur Physik, Johannes Gutenberg-Universit\"at
  Mainz, D55099 Mainz, Germany}
\author{Olga I. Vinogradova}
\affiliation{A.N. Frumkin Institute of Physical Chemistry and
  Electrochemistry, Russian Academy of Science, 31 Leninsky Prospect,
  119071 Moscow, Russia}
\affiliation{Department of Physics, M.V. Lomonosov Moscow State
  University, 119991 Moscow, Russia}
\affiliation{DWI, RWTH Aachen, Forckenbeckstra\ss e 50, 52056 Aachen,
  Germany}


\begin{abstract}

We study the effective slippage on superhydrophobic grooves with trapezoidal
cross-sections of various geometries (including the limiting cases of triangles and
rectangular stripes), by using two complementary approaches.  First,
dissipative particle dynamics (DPD) simulations of a flow past such surfaces
have been performed to validate an expression [E.~S. Asmolov and O.~I. Vinogradova,
J. Fluid Mech. \textbf{706}, 108 (2012)] that relates the eigenvalues of the
effective slip-length tensor for one-dimensional textures.  Second, we propose
theoretical estimates for the effective slip length
and calculate it numerically by solving the Stokes equation based on a
collocation method.
The comparison between the two approaches shows that they are in excellent
agreement.  Our results demonstrate that the effective slippage depends
strongly on the area-averaged slip, the amplitude of the roughness, and on the
fraction of solid in contact with the liquid.  To interpret these results, we
analyze flow singularities near slipping heterogeneities, and demonstrate that
they inhibit the effective slip and enhance the anisotropy of the flow.
Finally, we propose some guidelines to design optimal one-dimensional
superhydrophobic surfaces, motivated by potential applications
in microfluidics.
\end{abstract}


\maketitle

\section{Introduction}
\label{sec:introduction}

The design and fabrication of micro- and nanotextured surfaces has received
much attention during the past decade.  In case of a hydrophobic texture a
modified surface profile can lead to a very large contact angle, which induces
exceptional wetting properties~\cite{quere.d:2005}.  A remarkable mobility of
liquids on such superhydrophobic surfaces is observed in the Cassie state,
where the textures are filled with gas, which renders them ``self-cleaning''
and causes droplets to roll (rather than slide) under
gravity~\cite{richard.d:1999} and rebound (rather than spread) upon
impact~\cite{richard.d:2000,tsai.p:2010}.  Furthermore, patterned
superhydrophobic materials are important in context of fluid dynamics and their
superlubricating properties~\cite{vinogradova.oi:2012,mchale.g:2010}.  In
particular, superhydrophobic heterogeneous surfaces in the Cassie state exhibit
very low friction, and this drag reduction is associated with the large
slippage of liquids~\cite{bocquet2007,rothstein.jp:2010,vinogradova.oi:2010}.

It is very difficult to quantify the flow past heterogeneous surfaces.
However, analytical results can often be obtained using an effective
slip boundary condition, $\mathbf{b}_{\mathrm{eff}}$, at the imaginary smooth
homogeneous, but generally anisotropic surface~\cite{vinogradova.oi:2010,kamrin.k:2010}.
For anisotropic textures, the effective slip generally depends on the
direction of the flow and is a tensor, $\mathbf{b}_{\mathrm{%
eff}}\equiv \{b_{ij}^{\mathrm{eff}}\}$ represented by a symmetric, positive
definite $2\times 2$ matrix~\cite{Bazant2008}
\begin{equation}
\mathbf{b}_{\mathrm{eff}}=\mathbf{S}_{\theta }\left(
\begin{array}{cc}
b_{\mathrm{eff}}^{\parallel } & 0 \\
0 & b_{\mathrm{eff}}^{\perp }%
\end{array}%
\right) \mathbf{S}_{-\theta },  \label{beff_def1}
\end{equation}%
diagonalized by a rotation
\begin{equation}
\mathbf{S}_{\theta }=\left(
\begin{array}{cc}
\cos \theta & \sin \theta \\
-\sin \theta & \cos \theta%
\end{array}%
\right) .
\end{equation}%
Equation~(\ref{beff_def1}) allows us to calculate an effective slip in
any direction given by an angle $\theta $, provided the two eigenvalues of
the slip-length tensor, $b_{\mathrm{eff}%
}^{\parallel }$ ($\theta =0$) and $b_{\mathrm{eff}}^{\perp }$ ($\theta =\pi
/2$), are known.
The concept of an effective slip length tensor is general and can be applied
for arbitrary channel thickness~\cite{harting.j:2012}. It is a global
characteristic of a channel~\cite{vinogradova.oi:2010}, and the eigenvalues
usually depend not only on the parameters of the heterogeneous
surfaces, but also on the channel thickness.
However, for a thick channel (compared to a texture period, $L$) they
become a characteristic of a heterogeneous interface solely.

An important type of such surfaces are highly anisotropic textures, where a
(scalar) partial local slip $b(y)$ varies in only one direction.
Such surfaces can be created experimentally by making textured surfaces
with one dimensional surface profiles, and in the Cassie state, the local slip length will always reflect the relief of the texture. This is the consequence of the ``gas cushion model'', which takes into account that the
dissipation at the gas/liquid interface is dominated by the shearing of a
continuous gas layer~\cite{vinogradova.oi:1995a,andrienko.d:2003}
\begin{equation}
  \label{bgas}
  b(y) \simeq  \frac{\mu}{\mu_g} e(y),
\end{equation}
where $\mu_g$ and $\mu$ are dynamic viscosities of a gas and a liquid,
and $e$ is the thickness of the gas layer.
Eq.~(\ref{bgas}) represents an upper bound for a local slip at the gas area, which is attained in the limit of a small fraction of the solid phase. At a relatively large solid fraction~\cite{Ybert2007} or if the flux in the ``pockets'' of superhydrophobic surfaces is equal to zero~\cite{maynes1,nizkaya.tv:2013}, Eq.~(\ref{bgas}) could overestimate the local slip. However, even in all situations the local slip length profile, $b(y)$, necessarily follows the relief of the texture provided it is shallow enough, $
e\left( y\right) \ll L$~\cite{nizkaya.tv:2013}. Since in case of water Eq.~(\ref{bgas}) leads to $ b(y) \simeq  50 e(y)$, it is expected that the ``gas cushion model'' could be safely applied up to $b (y)/L = O(10)$ or so.

The flow properties on highly anisotropic surfaces can be highly peculiar.
In particular, off-diagonal effects may be obtained where a driving force
(such as pressure gradient, shear rate, or electric field) in one direction
induces a flow in a different direction, with a measurable and  useful
perpendicular component.This could be exploited for implementing transverse pumps,
mixers, flow detectors, and more~\cite{stroock2002b,ajdari2002}.
The main task is to find the connection between the eigenvalues of the slip-length
tensor and the parameters of such one-dimensional surface textures.
Up to now analytical results for the effective slip length have only been
obtained for the simplest geometry of rectangular grooves
\cite{lauga.e:2003,Belyaev2010,ng.co:2010,Zhou2012,harting.j:2012,asmolov:2013}.
We are unaware of any prior work that has obtained analytical results for other
types of one-dimensional textures.

However, some recently derived relations for the case of an anisotropic
isolated surface (corresponding to the thick channel limit) allow one to simplify
the analysis of more complex geometries, and to rationalize the limiting behavior of the
effective slip length in certain regimes.  We mention first that the transverse
component of the slip-length tensor was predicted to be exactly half the
longitudinal component that one would obtain if the local slip were multiplied
with a factor two, $2b(y)$~\cite{asmolov:2012}
\begin{equation}
b_{\mathrm{eff}}^{\bot }\left[ b\left( y\right) /L\right] =\frac{b_{\mathrm{%
eff}}^{\parallel }\left[ 2b\left( y\right) /L\right] }{2}.  \label{1D}
\end{equation}%
This relation can greatly simplify the analysis since it indicates that the
flow along any direction of the one-dimensional surface can be easily determined,
once the longitudinal component of the effective slip tensor is found.
Equation (\ref{1D}) has been recently verified for a texture with sinusoidal
superhydrophobic grooves by using a Lattice Boltzmann (LB) method \cite{asmolov:2013b},
and for weakly slipping stripes by using dissipative particle dynamics (DPD)
simulations~\cite{asmolov:2013}.

Another useful relation addresses the particular case where the local slip is large,
$b(y)/L \gg 1$. In that case, the effective  slip length tensor $\textbf{b}_{\rm eff}$
becomes isotropic and is given by \cite{asmolov:2013b,hendy.sc:2007}
\begin{equation}\label{eq:b_large}
b^{\parallel}_{\rm eff} \simeq b^{\perp}_{\rm eff} \simeq \left\langle \frac{1}{b (y)} \right\rangle^{-1},
\end{equation}
where $\left\langle 1/b (y) \right\rangle$ is the mean inverse local slip.
Finally, for weakly slipping patterns with $b(y)/L \ll 1$, the flow again becomes
isotropic, and its value is equal to the surface average of the local slip
length~\cite{Belyaev2010,kamrin.k:2010}:
\begin{equation}\label{eq:b_small}
b^{\parallel}_{\rm eff} \simeq b^{\perp}_{\rm eff} \simeq \langle b (y) \rangle = \langle b \rangle,
\end{equation}
Note that Eq.~(\ref{eq:b_small}) is expected to be accurate for a
continuous local slip profile only. If the local slip length exhibits step-like jumps at
the edges of heterogeneities, the expansions leading to Eq.~(\ref{eq:b_small})
become questionable. Higher order contributions to the effective
slip become comparable to the first-order term given by Eq.~(\ref{eq:b_small})
and can no longer be ignored \cite{asmolov:2013}.

Computer simulations and numerical modeling can shed light on
flow phenomena past one-dimensional surfaces. In efforts to better understand the
connection between the parameters of the texture and the effective slip lengths,
several groups have performed continuum and Molecular Dynamics simulations of flow
past anisotropic surfaces. Most of these studies focused on calculating eigenvalues
of the effective slip-length tensor for superhydrophobic
stripes~\cite{cottin.c:2004,Priezjev2005,priezjev.n:2011}.
To study various aspects of a flow past striped walls, recent works employed
Dissipative particle dynamics (DPD)~\cite{Zhou2012,asmolov:2013,Tretyakov2013}
and Lattice Boltzmann (LB)~\cite{harting.j:2012} methods.
In the case of stripes with a piecewise constant local slip profile (i.e.,
the profile features steplike jumps at the boundaries of the stripes),
the singularity in the slip profile induces singularities both in the pressure
and velocity gradient, which generates an additional mechanism for dissipation
even at small local slip~\cite{asmolov:2012,asmolov:2013}.
Anisotropic one-dimensional texture with continuously varying patterns can
produce a very large effective slip.  This has been confirmed in recent
LB simulations studies of sinusoidal textures~\cite{asmolov:2013b}.
One-dimensional continuous surfaces may also exhibit singularities
(for example, the first derivative may be discontinuous). However, we are
unaware of any previous work that has addressed the question of effective slip
lengths past such textures.

\begin{figure}[htbp]
  \centering
  \includegraphics[width=1.0\columnwidth]{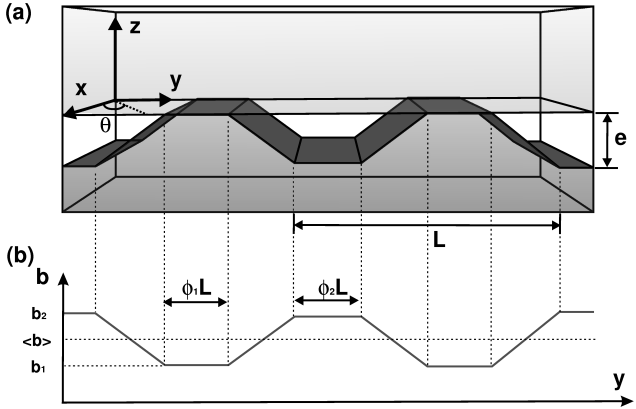}
  \caption{Sketch of the trapezoidal superhydrophobic surface in the Cassie state (a),
  and, according to Eq.~(\ref{bgas}), of the corresponding local slip length (b).}
  \label{fig:sketch}
\end{figure}

In this paper we study the friction properties on trapezoidal superhydrophobic
grooves [see Fig. \ref{fig:sketch}(a)].  This texture is more general than
rectangular grooves since it provides an additional geometrical structure
parameter, i.e., the trapezoid base width.  Also, the trapezoid texture can be
extended to its extremes, such as a triangular patterns, where the liquid/solid
contact area vanishes, and rectangular reliefs.  Superhydrophobic trapezoidal
textures were shown to give large contact angles, and to provide stable Cassie
states~\cite{li.w:2010}.  An obvious advantage of the trapezoidal surface relief
is that the manufactured texture becomes mechanically more stable against
bending compared to rectangular grooves.  Furthermore, it can be
naturally formed from dilute rectangular grooves as a result of
elasto-capillarity~\cite{tanaka.t:1993}.  This type of surfaces has already
been intensively used in slippage experiments~\cite{choi.ch:2006}.  However,
the impact of trapezoidal texture reliefs on transport and hydrodynamic
slippage remains largely unknown.

Our study is based on the use of DPD simulations, which we have put forward
in our previous papers~\cite{Zhou2012,asmolov:2013}, on the numerical solution of
the Stokes equations (a collocation method), and on theoretical analysis.  In
particular, we validate Eq.~(\ref{1D}) for trapezoidal textures with various
parameters, we apply Eq.~(\ref{eq:b_large}) to evaluate an effective slip for
strongly slipping  patterns, we discuss the singularities in velocities caused by
inflection points in the local slip profiles, which in particular lead to
deviations from Eq.~(\ref{eq:b_small}), and we finally suggest the optimal design
of trapezoidal patterns.

\section{Model and Governing Equations}
\label{sec:theory}

We consider creeping flow along a plane anisotropic superhydrophobic wall, using a Cartesian
coordinate system $(x,y,z)$ (Fig. \ref{fig:sketch}).  As in most previous
publications~\cite{bocquet2007,belyaev.av:2010b,Priezjev2005,lauga.e:2003,Ybert2007,asmolov_etal:2011,feuillebois.f:2009, harting.j:2012}, we model
the superhydrophobic plate as a flat heterogeneous interface. In such a description,  we neglect an additional mechanism for a dissipation connected with the meniscus curvature~\cite{sbragaglia.m:2007,harting.j:2008,Ybert2007}. Note however that such an ideal situation has been achieved in many recent experiments~\cite{steinberger.a:2007,karatay.e:2013,haase.as:2013}. The origin of
coordinates is placed at the flat interface, characterized by a slip length
$b(y)$, spatially varying in one direction, and the texture varies over a
period $L$.

The local slip length is taken to be a piecewise linear function.
It includes two regions with a constant slip: one where the slip length assumes
its minimum value, $b_1$, which has area fraction of $\phi_1$, and another with
area fraction $\phi_2$ where the slip length is maximal. These regions are connected
by two transition regions with equal surface fractions $(1-\phi_1-\phi_2)/2$, where
the local slip length varies linearly from $b_1$ to $b_2$. If we rewrite the
coordinate $y$ in the form $y=(n+t)L$, where $n$ is an integer number and
$|t|\le \frac{1}{2}$, the local slip-length profile can be expressed in terms of $t$:
\begin{equation}
  \label{eq:profile}
  b(y) =
  \begin{cases}
    b_{1} , & |t| \le \frac{\phi_1}{2} \\
    b_{1} + \frac{ 2 (b_2 - b_1) }{1-\phi_1-\phi_2} \big( |t| - \frac{\phi_1}{2} \big), & \frac{\phi_1}{2} < |t| \le \frac{1-\phi_2}{2} \\
    b_{2}, & \frac{1-\phi_2}{2} < |t| \le \frac{1}{2}
  \end{cases}
\end{equation}
For $\phi_1=\phi_2=0$, we get a triangular profile
\begin{equation}
  b(y) = b_{1} + 2(b_2-b_1) |t|, \quad |t| \le \frac{1}{2} .
\end{equation}
Note that regular trapezoidal and triangular profiles are continuous functions
(albeit with a discontinuous first derivative).
However, in the limit $\phi_1+\phi_2=1$, the profile reduces to the discontinuous
rectangular slip length profile (alternating stripe profile) studied in earlier work
\begin{equation}
  b(y)=
  \begin{cases}
    b_{1}, & |t| \le \frac{\phi_1}{2} \\
    b_{2}, & \frac{\phi_1}{2} < |t| \le \frac{1}{2}
  \end{cases}
\end{equation}

Our results apply to a single surface in a thick channel, where effective
hydrodynamic slip is determined by flow at the scale of roughness, so that the velocity profile
sufficiently far above the surface may be considered as a linear shear flow~\cite{bocquet2007,vinogradova.oi:2010}. However it does not apply
to a thin~\cite{feuillebois.f:2009} or an arbitrary~\cite{harting.j:2012} channel situation, where the effective slip scales with the channel width. Note that the flow in thin channels with superhydrophobic trapezoidal textures has been considered in recent work~\cite{nizkaya.tv:2013}.

Dimensionless variables are defined by using $L$ as a reference length scale,
the inverse shear rate $1/G$ sufficiently far from the surface as a reference
time scale, and setting the fluid kinematic viscosity, $\nu$, to unity (this
sets the scale of mass). At small Reynolds number, $%
Re=GL^{2}/\nu,$
the dimensionless velocity and pressure fields, $\mathbf{v}$ and $p$, satisfy Stokes equations

\begin{gather}
\mathbf{\nabla }\cdot \mathbf{v}=0,  \label{Se} \\
\mathbf{\nabla }p-\Delta \mathbf{v}=\mathbf{0}.  \notag
\end{gather}
Due to Eq.~(\ref{1D}), it is sufficient to consider the flow parallel to the
pattern ($x$-axis). In this case, the velocity has only one component in the
$x$-direction $\mathbf{v} = (v(y,z),0,0)$, and we seek a solution for the velocity profile
of the form
\begin{equation}
  v=U+u,
\end{equation}
where $U=z$ is the undisturbed linear shear flow.
The perturbation of the flow,
 $u(y,z)$, which is caused by the presence of the
texture and decays far from the interface. It follows from Stokes' equations that $\partial_y p=\partial_z p=0$. Since the pressure is constant
far from the superhydrophobic surface this implies that it also remains constant in the entire flow region. The velocity perturbation satisfies a dimensionless Laplace equation%
\begin{equation}
  \label{eq:laplace}
  \Delta u = 0.
\end{equation}
The boundary conditions at the wall and at infinity
are defined in the usual way%
\begin{eqnarray}
  \label{eq:bc_z0}
  z=0: & & u - \frac{b(y)}{L} \partial_z u = \frac{b(y)}{L}, \\
  \label{eq:bc_zinfty}
  z\rightarrow \infty: & & \partial_z u = 0.
\end{eqnarray}

The general solution to Eq.~(\ref{eq:laplace}), decaying at infinity,
can be represented in terms of a cosine Fourier series as
\begin{equation}
  u = \frac{a^0}{2} + \sum_{n=1}^{\infty} a^n \exp(-2\pi n z) \cos(2\pi n y),
\end{equation}
where $a^n$ are coefficients to be found by applying the appropriate
boundary condition (\ref{eq:bc_z0}).
It can be written in terms of the Fourier coefficients as%
\begin{equation}
\frac{a^{0}}{2}+\sum_{n=1}^{\infty }\left[ 1+2\pi n\frac{b(y)}{L}\left(
y\right) \right] a^{n}\cos \left( 2\pi ny\right) =\frac{b(y)}{L}
\label{bc_ft}
\end{equation}%
Equation~(\ref{bc_ft}) was used to determine the longitudinal
effective slip numerically. This was done using a method similar
to~\cite{asmolov:2013}. We truncate the sum in Eq.~(\ref{bc_ft})
at some cut-off number $N$ (usually $N=1001$) and evaluate it
in the points $y_{l}=l/2\left( N-1\right) ,$ where $l$ are
numbers varying from $0$ to $N-1$. The problem is then reduced to a
linear system of equations, $A_{n}^{l}a^{n} L= b^{l},$ where
$A_{0}^{l}=1/2$; $A_{n}^{l} =\left[ 1+2\pi n b\left(
y_{l}\right)/L \right] \cos \left( 2\pi ny_{l}\right) $,
$n>0$ and $b^{l}=b \left( y_{l}\right) .$ The system is solved
using the IMSL routine LSARG. The transverse effective slip was then
calculated using Eq.~(\ref{1D}).

\section{Dissipative Particle Dynamics Simulation}
\label{sec:simulation}

Mesoscale fluid simulations were performed using the dissipative
particle dynamics (DPD) method \cite{Hoogerbrugge1992, Espanol1995, Groot1997}.
More specifically, we use a version of DPD without conservative
interactions \cite{Soddemann2003} and combine that with a tunable-slip
method \cite{Smiatek2008} to model patterned surfaces.
The tunable-slip method allows one to implement arbitrary hydrodynamic
boundary condition in coarse-grained simulations.
In the following, we give a brief description of the model and
introduce the implementation of planar surfaces with
arbitrary one-dimensional slip profile.

The basic DPD equations involve pair interactions between fluid particles.
The pair forces are equal in magnitude but opposite in direction; thus
the momentum is conserved for the whole system.
This leads to the correct long-time hydrodynamic behavior (i.e. Navier-Stokes).
The pair force consists of a dissipative contribution and a random component
\begin{equation}
  \mathbf{F}_{ij}^{\rm DPD} = \mathbf{F}_{ij}^{\rm D} + \mathbf{F}_{ij}^{\rm R}.
\end{equation}
The dissipative part $\mathbf{F}_{ij}^{\rm D}$ is proportional to the
relative velocity between two particles
$\mathbf{v}_{ij}=\mathbf{v}_i-\mathbf{v}_j$,
\begin{equation}
  \label{eq:dpd_F_D}
  \mathbf{F}_{ij}^{\rm D} = - \gamma_{\rm DPD} \, \omega_{\rm D}(r_{ij})
  (\mathbf{v}_{ij} \cdot \hat{\mathbf{r}}_{ij}) \hat{\mathbf{r}}_{ij} ,
\end{equation}
with a local distance-dependent friction coefficient $\gamma_{\rm DPD}
\omega_{\rm D}(r_{ij})$.
The weight function $\omega_{\rm D}(r_{ij})$ is a monotonically decreasing
function of $r_{ij}$ and vanishes at a given cutoff, and
$\gamma_{\rm DPD}$ is a parameter that controls the strength of the
dissipation.
The random force $\mathbf{F}_{ij}^{\rm R}$ has the form
\begin{equation}
  \mathbf{F}_{ij}^{\rm R} = \sqrt{ 2 k_B T \gamma_{\rm DPD} \, \omega_{\rm D}(r_{ij})
  } \xi_{ij} \hat{\mathbf{r}}_{ij},
\end{equation}
where $\xi_{ij}=\xi_{ji}$ is a symmetric random number with a zero mean
and a unit variance.
The amplitude of the stochastic contribution is related to the
dissipative contribution by a fluctuation-dissipation theorem in order
to ensure that the model reproduces the correct equilibrium distribution.

The tunable-slip method \cite{Smiatek2008} introduces the wall
interaction in a similar fashion.
The force exerted by the wall on particle $i$ is given by
\begin{equation}
  \mathbf{F}_i^{\rm wall} = \mathbf{F}_i^{\rm WCA}+\mathbf{F}_i^{\rm D} + \mathbf{F}_i^{\rm R} .
\end{equation}
The first term is a repulsive interaction that prevents the fluid particles
from penetrating the wall.
It can be written in term of the gradient of a Weeks-Chandler-Andersen
potential \cite{WCA}:
\begin{eqnarray}
  && \mathbf{F}_i^{\rm WCA} = - \nabla V(z), \nonumber \\
  && V(z) = \left\{ \begin{array}{cl} 4 \epsilon [ (\frac{\sigma}{z})^{12}
  - (\frac{\sigma}{z})^6 + \frac{1}{4} ]  \quad & z < \sqrt[6]{2} \, \sigma \\
0 & z \ge \sqrt[6]{2} \, \sigma \end{array} \right.
\end{eqnarray}
where $z$ is the distance between the fluid particle and the wall,
assuming the wall lies in $xy$ plane, and $\epsilon$ and $\sigma$ set
the energy and length scales.
In the following, physical quantities will be reported in a model unit
system of $\sigma$ (length), $m$ (mass), and $\epsilon$ (energy).
The dissipative contribution is similar to Eq.~(\ref{eq:dpd_F_D}),
with the velocity replaced by the relative particle velocity with
respect to the wall. For immobile walls, the dissipative part can
be written as
\begin{equation}
  \mathbf{F}_i^{\rm D} = - \gamma_{\rm L} \omega_{\rm L}(z) \mathbf{v}_i .
\end{equation}
The parameter $\gamma_{\rm L}$ characterizes the strength of the wall
friction and can be used to tune the value of the slip length.
For example, $\gamma_{\rm L}=0$ results in a perfectly slippery wall,
whereas a positive value of $\gamma_{\rm L}$ leads to a finite slip length.
The random force has the form
\begin{equation}
  \mathbf{F}_i^{\rm R} = \sqrt{2 k_B T \gamma_{\rm L} \omega_{\rm L}(z)} \boldsymbol{\xi}_i ,
\end{equation}
where each component of $\boldsymbol{\xi}_i$ is a random variable.
Again, the magnitude of the random force is chosen such that the
fluctuation-dissipation theorem is satisfied.

\begin{table*}
\centering
\caption{Parameters used in the DPD simulations.}
\begin{tabular}{l p{0.4cm} l}
\hline
fluid density $\rho$ & & $3.75 \sigma^{-3}$ \\
friction coefficient for DPD interaction $\gamma_{\rm DPD}$ & &
$5.0\sqrt{m\epsilon}/\sigma$ \\
cutoff for DPD interaction & & 1.0 $\sigma$\\
no-slip boundary condition $\gamma_{\rm L}$ ($b=0$) & & $5.26 \sqrt{m\epsilon}/\sigma$ \\
cutoff for wall interaction & & 2.0 $\sigma$ \\
dynamic viscosity $\mu$ & & $1.35 \pm 0.01 \sqrt{m\epsilon}/\sigma^2$ \\
hydrodynamic boundary from the simulation wall & & $1.06 \pm 0.12
\sigma$ \\
simulation box & & $20 \sigma \times 40 \sigma \times 82 \sigma$ \\
pattern period, $L$ & & $40 \sigma$ \\
\hline
\end{tabular}
\label{tab:parameters}
\end{table*}

\begin{figure}[htbp]
  \centering
  \includegraphics[width=1.0\columnwidth]{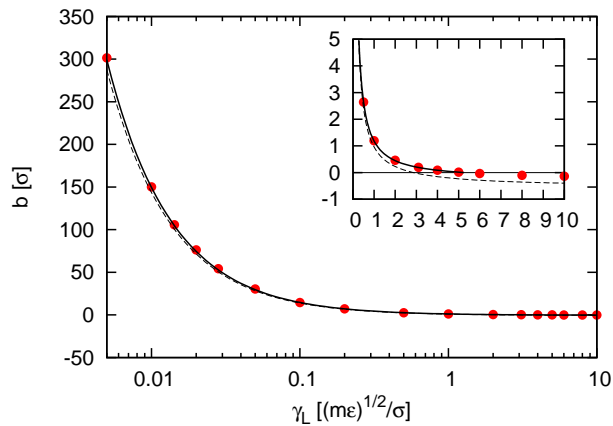}
  \caption{The relation between the slip length $b$ and the wall
    friction coefficient $\gamma_{\rm L}$, for a wall interaction cutoff $2.0
    \sigma$. The inset shows an enlarged portion of the region where
    the slip length is close to zero. The no-slip boundary condition
    is implemented by using $\gamma_{\rm L} = 5.26
    \sqrt{m\epsilon}/\sigma$. The solid lines are the fitting result
    Eq. (\ref{eq:b_fit}), and the dashed curves show the analytical
    prediction.}
  \label{fig:b_gammaL}
\end{figure}

For homogeneous surfaces, a useful analytical expression can be derived
which establishes a relation between the slip length $b$ and the simulation
parameters (the wall friction parameter $\gamma_{\rm L}$ and the cutoff $r_c$)
to a good approximation \cite{Smiatek2008}.  However, the
accuracy of the analytic result is not satisfactory at small slip lengths.  For
the purpose of the present work, we need an accurate expression for the relation
between the slip length and the tuning parameter $\gamma_{\rm L}$.
Therefore, we have simulated plane Poiseuille and Couette flows with different
$\gamma_{\rm L}$ to obtain the slip length and the position of the hydrodynamic
boundary.  The simulation results were fitted to a truncated Laurent series
up to second order in $\gamma_{\rm L}$ and first order in $1/\gamma_{\rm L}$.
The simulation data and the fitted curve are shown in Fig.~\ref{fig:b_gammaL}.
The fit yields the expression
\begin{equation}
  \label{eq:b_fit}
  b = \frac{1.498}{\gamma_{\rm L}} - 0.3291 + 0.01457 \gamma_{\rm L} - 0.001101
  \gamma_{\rm L}^2.
\end{equation}
This formula relates the wall friction coefficient $\gamma_{\rm L}$, which
is set in the simulation, to the slip length $b$.
Also shown in Fig.~\ref{fig:b_gammaL} is the analytic result in dashed
line, which agrees with the simulation data quite well at large slip
length.

One important observation is that both the slip length $b$ and the friction
coefficient $\gamma_{\rm L}$ are local quantities.
Thus, it is reasonable to assume that the relation (\ref{eq:b_fit}) also
holds for inhomogeneous surfaces, as long as the slip length varies on
length scales much larger than the range of the weight functions,
$\omega_{\rm L} (z)$ and $\omega_{\rm D}(r)$.
For a patterned surface with $L$ much larger than $\sigma$, the local friction coefficient $\gamma_{\rm L}(y)$ can be calculated from the slip profile $b(y)$ by inverting Eq.~(\ref{eq:b_fit}).
The resulting profile $\gamma_{\rm L}(y)$ is then used in simulations
to set up the patterned surfaces.

The simulations are carried out using the open source simulation package ESPResSo \cite{ESPResSo}.
Modifications have been made to incorporate the patterned surfaces.
Table \ref{tab:parameters} summarizes the simulation parameters.
In simulations, the patterned surface has a period of $L=40\sigma$.
The simulation box is a rectangular cuboid of size $20\sigma \times 40
\sigma \times 82\sigma$,
periodic boundary conditions are assumed in the $xy$ plane, and the
two symmetric patterned surfaces are located at $z=0 \sigma$ and $z=82
\sigma$.
The separation between the two surfaces is about twice the pattern
periodicity.
We have checked that increasing the channel thickness does not change the
results.

The simulation starts with randomly distributed particles inside the channel,
and flow is induced by applying a body force to all particles. Body forces
used in this simulation were chosen small enough that the flow velocity
near the wall was less than $0.1\,\sqrt{\epsilon/m}$, in order to ensure
that the flow is strictly laminar and effects of finite Reynolds numbers
do not affect the simulation results.
The magnitude of the slip velocity also affects the spatial resolution in modeling the variation of the slip length.
Since the trajectories of the fluid particles are integrated at discrete time steps, the friction experienced by the particle between time steps is determined from the early instant.
Therefore, we used a small time step of $0.01 \sqrt{m/\epsilon} \sigma$.
Due to the large system size, the relaxation time
required for reaching a steady state was very large as well (over $10^6$ time
steps). The flow velocity is small in comparison with the thermal fluctuation;
thus many time steps are necessary to gather sufficient statistics.  In this
work, velocity profiles were averaged over $4\times10^5$ time steps.  A fit to
the plane Poiseuille flow then gives the effective slip length $b_{\rm eff}$
\cite{Zhou2012}.  The error bars are obtained by six independent runs with
different initialization.

Finally, we note that in case of a triangular local slip, only a very small
region near the no-slip point contributes to the shear stress  at large
$b_{2}$.  This presents a complication for simulations since fluid particles with a
finite translational velocity travel a small distance over the no-slip region
before reaching the zero averaged velocity.  This distance is related to the
initial velocity of the fluid particles when they enter the no-slip region,
which is in general proportional to $b_2$ and the pressure gradient (the body
force in simulation).  To obtain accurate simulation data at large
$b_2/L$, one has to reduce the body force significantly, and the necessary
simulation times increase prohibitively. Therefore, only simulation results
for $b_{2}/L = O (1)$ and smaller are presented for this texture.

\section{Eigenvalues of the slip-length tensor}

Here we present the results for the eigenvalues of the slip-length tensor
obtained by simulations of different local slip profiles.  We first verify
Eq.~(\ref{1D}), derived earlier by some of us~\cite{asmolov:2012} to relate
the longitudinal and transverse effective slip lengths. We use local slip
profiles with $\phi_1=\phi_2$. In that special case, the profile $b(y)$ can be
characterized by three parameters: the area fraction $\phi_1$ of the no-slip region,
the surface-averaged slip length $\langle b \rangle =(b_1+b_2)/2$, and the heterogeneity factor,
$\alpha$, which is a measure for the relative variation of the slip length
and defined as
\begin{equation}
  \alpha=\frac{b_2-b_1}{2 \langle b \rangle}=\frac{b_2-b_1}{b_2+b_1}
\end{equation}
A homogeneous surface then corresponds to $\alpha=0$, and low values of $\alpha$
corresponds to small $b_2/b_1$. Patterns involving no-slip regions,  (i.e.
$b_1=0$) have $\alpha=1$. Such a no-slip assumption can be justified provided
$b_2/b_1 \gg 1$, which is typical for many situations. Indeed, for most smooth
hydrophobic surfaces, the slip length is of the order of a few tens of
nm~\cite{vinogradova:03,vinogradova.oi:2009,charlaix.e:2005,joly.l:2006}, which
is very small compared to the slip length values of up to tens or even hundreds
of $\mu$m slip lengths that may be obtained on the gas areas.

\subsection{Fixed average slip}

\begin{figure}[htbp]
  \centering
  \includegraphics[width=1.0\columnwidth]{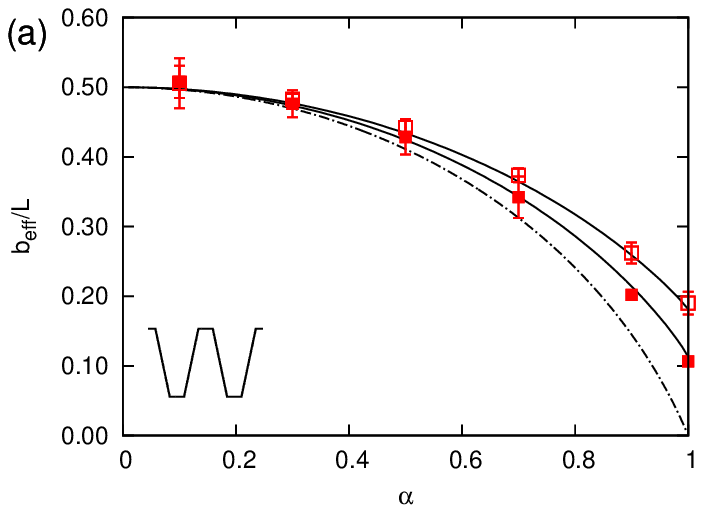}
  \includegraphics[width=1.0\columnwidth]{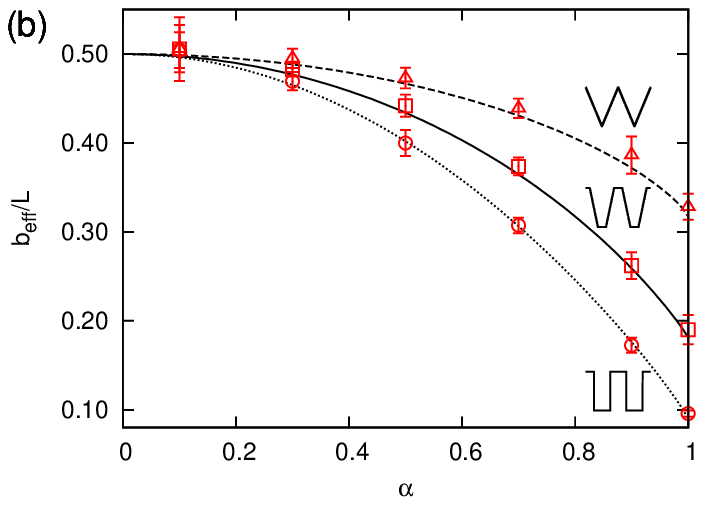}
  \caption{(a) Longitudinal (open symbols) and transverse (closed symbols)
  effective slip for a trapezoidal texture with an average slip $\langle b \rangle/L=0.5$.
  Solid curves show corresponding numerical solutions. Dash-dotted curve shows
  asymptotic solution in the limit of the large local slip [Eq.~(\ref{eq:b_large2})].
  (b) Longitudinal effective slip length simulated (from top to bottom) for
  triangular, trapezoidal, and rectangular profiles with the same average
  slip $\langle b \rangle/L=0.5$ (symbols) and corresponding numerical data.  }
  \label{fig:alpha}
\end{figure}

Let us first investigate the effect of $\alpha$ on the effective slip at fixed
$\langle b \rangle/L=0.5$.  Figure \ref{fig:alpha}(a) shows the simulation results for
eigenvalues of the slip length tensor as a function of $\alpha$ obtained for a
trapezoidal local slip ($\phi_1=0.25$).  Also included are numerical results
(solid curves) calculated with Eqs.~(\ref{1D}) and (\ref{bc_ft}).  The
agreement between the simulation and numerical results is quite good for all
$\alpha$.  This demonstrates both the accuracy of our simulations, and the
validity of Eq.~(\ref{1D}).  The data presented in Fig.~\ref{fig:alpha}(a) show
that the longitudinal slip is larger than the transverse slip, i.e., the flow
is anisotropic, and the anisotropy increases with $\alpha$.  The effective slip
assumes its maximum value at $\alpha=0$, when the surface is homogeneous, and then
decreases monotonously. This implies that the effective slip is found to be
always smaller than the surface-averaged value, and the difference from the
average value increases with the heterogeneity. A similar behavior was
reported in prior work for other types of
patterns~\cite{vinogradova1999,asmolov:2013b}. In the case of large local
slip, the effective slip length can be obtained from Eq.~(\ref{eq:b_large}),
which gives for a trapezoidal slip profile, Eq.~(\ref{eq:profile}):
\begin{equation}
  \label{eq:b_large2}
  b_{\mathrm{eff}}\simeq\left[ \frac{\phi _{1}}{b_{1}}+\frac{\phi _{2}}{b_{2}}+%
\frac{1-\phi _{1}-\phi _{2}}{2\left( b_{2}-b_{1}\right) }\ln \left( \frac{%
b_{2}}{b_{1}}\right) \right] ^{-1}.
\end{equation}

Figure \ref{fig:alpha}(a) includes the theoretical curve expected for
this (isotropic) case (dash-dotted line). We note that the slip length
is always smaller than the average slip, $\langle b \rangle/L=0.5$.
Perhaps the most interesting and important result here is that
Eq.~(\ref{eq:b_large2}) remains surprisingly accurate even in the case of
moderate, but not vanishing, local slip, provided $\alpha$ is small enough.
We remind the reader that Eq.~(\ref{eq:b_large2}) may not be used when
$\alpha=1$, i.e., for textures with no-slip regions, $b_1=0$, where it
would simply predicts the effective slip length to be zero.

Qualitatively similar results were obtained for rectangular stripes
and triangular local slip profiles. Therefore we do not show these
results here. We only recall that in the case of stripes, the
asymptotic result for large slip gives
(see~\cite{asmolov:2013b,cottin.c:2004} for the original derivations),
\begin{equation}
  \label{eq:b_large_stripes}
  b_{\mathrm{eff}}\simeq\left[ \frac{\phi _{1}}{b_{1}}+\frac{\phi _{2}}{b_{2}} \right] ^{-1},
\end{equation}
and for triangular profiles, one can derive
\begin{equation}
  \label{eq:b_large_triangle}
  b_{\mathrm{eff}}\simeq \frac{2 (b_2 - b_1)}{\ln (b_2/b_1)}.
\end{equation}

To examine the difference between the different slip profiles, the results for a
longitudinal effective slip for a texture with a trapezoidal local slip from
Fig.~\ref{fig:alpha}(a) are reproduced in Fig.~\ref{fig:alpha}(b).  Also
included in Fig.~\ref{fig:alpha}(b) are numerical and simulation data for
rectangular ($\phi_1=0.5$) and triangular ($\phi_1=0$) profiles with the same
average slip.  The data show that the triangular profile gives a larger, and
the rectangular one a smaller effective slip than that resulting from the
trapezoidal local slip. We note that $\phi_1$ is very different for these
slip-length profiles. This suggests that one of the key parameters
determining the effective slip is the area fraction of solid in contact with the
liquid.  If it is very small, the effective slip length becomes large.
Another important factor is the presence of flow singularities near slipping
heterogeneities as we will discuss below.

\subsection{Fixed heterogeneity factor}

\begin{figure}[htbp]
  \centering
  \includegraphics[width=1.0\columnwidth]{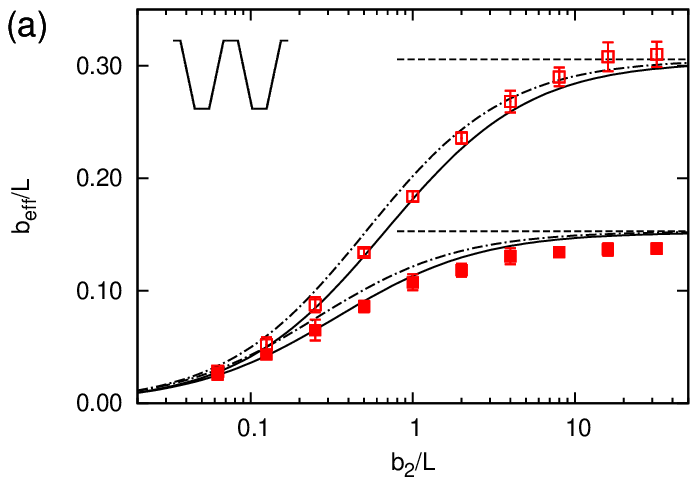}
  \includegraphics[width=1.0\columnwidth]{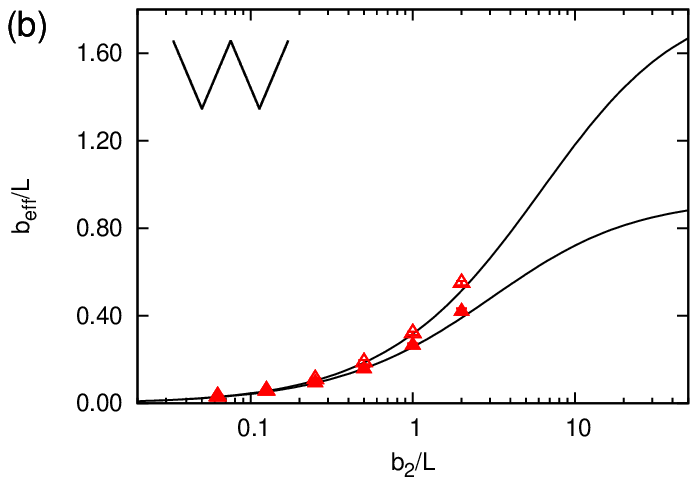}
   \caption{Effective slip lengths as a function of $b_{2}/L$ at $\alpha=1$ for
   (a) trapezoidal and (b) triangular local slip. Symbols show simulation data for
   longitudinal (open) and transverse (closed) effective slip lengths.
   The solid curves are results of numerical solutions.  Dash-dotted curves
   show the predictions of analytic formulae [Eqs.~(\ref{eq:b_par_bv}, \ref{eq:b_per_bv})]
   for stripes with the same fraction of no-slip areas. Dashed lines show the asymptotic
   values for stripes at large $b_2/L$ [Eq.~(\ref{eq:b_infty1})].}
  \label{fig:slip}
\end{figure}

We now fix $\alpha=1$ and consider flows that are maximally anisotropic for
a given pattern, which  corresponds to $b_{1}=0$ (no-slip areas) and $b_{2}=2\langle b \rangle$.
It is instructive to focus first on the role of $b_2$. Fig.~\ref{fig:slip}
shows the eigenvalues of the slip-length tensor as a function of $b_{2}/L$
as obtained from simulations (symbols). As before we include numerical results
(solid curves) calculated with Eqs.~(\ref{1D}) and (\ref{bc_ft}). We remind that the ``gas cushion model'', Eq.~(\ref{bgas}), becomes very approximate at $b_2/L \gg 1$, and the local slip profile could not necessarily follow the relief of the texture. It is, however, instructive to include large values of $b_2/L$ into our consideration below since this will provide us with some guidance.

The eigenvalues of the slip-length tensor for a superhydrophobic trapezoidal
texture are shown in Fig.~\ref{fig:slip}(a).  The agreement between the
theoretical and simulation data is very good for all $b_2/L$, again confirming
the accuracy of the simulations and the validity of Eq.~(\ref{1D}). Qualitatively,
the results are similar to those obtained earlier for a striped
surfaces~\cite{Belyaev2010}: At very small $b_2/L$, the effective slip
is small and appears to be isotropic, in accordance to predictions of
Eq.~(\ref{eq:b_small}).  It becomes anisotropic and increases for larger
$b_2/L$, and saturates at $b_2/L \gg 1$.  In case of stripes with alternating
regions of no-slip ($b_{1}=0$) and partial-slip ($b_{2}=2\langle b \rangle$) areas the
situation can be described by simple analytic formulae~\cite{Belyaev2010}
\begin{eqnarray}
  \label{eq:b_par_bv}
  \frac{b_{\rm eff}^{\parallel}}{L} &\simeq& \frac{1}{\pi} \frac{\displaystyle \ln
    \left[ \sec(\frac{\pi \phi_2}{2}) \right] }{\displaystyle 1 +
    \frac{L}{\pi b_{2}} \ln \left[\sec(\frac{\pi \phi_2}{2}) +
      \tan(\frac{\pi \phi_2}{2}) \right]} ,\\
  \label{eq:b_per_bv}
  \frac{b_{\rm eff}^{\perp}}{L} &\simeq& \frac{1}{2\pi} \frac{\displaystyle \ln
    \left[ \sec(\frac{\pi \phi_2}{2}) \right] }{\displaystyle 1 +
    \frac{L}{2\pi b_{2}} \ln \left[\sec(\frac{\pi \phi_2}{2}) +
      \tan(\frac{\pi \phi_2}{2}) \right]}
\end{eqnarray}
In the limit of $b_{2} \gg L$, these expressions turn into those derived earlier
for perfect slip stripes~\cite{philip.jr:1972,lauga.e:2003}
\begin{equation}
  \label{eq:b_infty1}
  \frac{b_{\rm eff}^{\parallel}}{L} = \frac{2b_{\rm eff}^{\perp}}{L}
  \simeq \frac{1}{\pi} \ln \left[ \sec \left(
      \frac{\pi\phi_2}{2} \right) \right] .
\end{equation}

For comparison, we plot the above expressions (using the value $\phi_1=0.25$
of the trapezoidal texture) in Fig.~\ref{fig:slip}(a) as dash-dotted
[Eqs.~(\ref{eq:b_par_bv},\ref{eq:b_per_bv})] and dashed
[Eq.~(\ref{eq:b_infty1}] lines.  One can see that with these parameters, the
results for stripes practically coincides with those for trapezoidal slip at
very small and very large  $b_{2}/L$.  At intermediate local slip there is a
small discrepancy, suggesting that the equations for partial slip stripes
overestimate the effective slip for trapezoidal texture. We will return to a
discussion of these results later in the text.

Figure \ref{fig:slip}(b) plots the simulation results for a triangular local
slip profile, which is particularly interesting since $\phi_1 =0$. As explained above,
simulation data could only be obtained for small values of $b_2$ in this
case. In contrast, our numerical approach allowed us to calculate the effective
slip length over a much larger interval of $b_2/L$. These numerically computed slip lengths are also included in
Fig.~\ref{fig:slip}(b). In the regime where simulation data are
available, the agreement with the simulation and numerical
results is excellent, therefore validating Eq.(\ref{1D}). It is interesting
to note that according to the numerical calculations, the effective slip lengths
grow weakly with $b_2/L$ at large $b_2/L$ and likely do not saturate, in contrast
to rectangular and trapezoidal textures.

\begin{figure}[htbp]
  \centering
  \includegraphics[width=1.0\columnwidth]{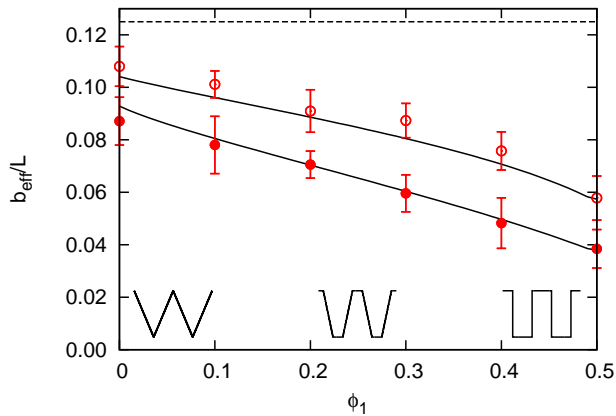}
  \caption{Effective slip lengths as functions of $\phi_1$ for surfaces with $\alpha=1$ and $\langle b \rangle/L=0.125$.
  The open and closed symbols are simulation results for the longitudinal and transverse
  effective slip lengths, respectively. Solid lines are numerical results.
  Dashed line shows predictions of Eq.~(\ref{eq:b_small}).}
  \label{fig:phi}
\end{figure}

Finally we compare the effective slip of surfaces with the same
$\alpha$ and $\langle b \rangle/L$, but different $\phi_1$.
Fig.~\ref{fig:phi} shows the corresponding curves for a maximal heterogeneity
factor, ($\alpha=1$), and a relatively small $\langle b \rangle/L=0.125$.
The results for larger $\langle b \rangle/L$ are qualitatively similar and not shown here.
Only the values of the effective slip lengths become larger.) We note that
by varying $\phi_1$ from 0 to 0.5, we change the local slip profile from triangular
($\phi_1=0$) to rectangular (stripes, $\phi_1=0.5$).
Regular trapezoidal textures correspond to intermediate values of $\phi_1$.
Fig.~\ref{fig:phi} also shows the analytical prediction of
Eq.~(\ref{eq:b_small}) for weakly slipping surfaces, i.e., an effective
slip $b_{\rm eff}^{\parallel,{\perp} } \simeq \langle b \rangle$ which is isotropic
and independent on $\phi_1$. The effective slip according to simulations
and numerical results is always smaller than the first-order theoretical prediction,
and the deviations from the area-average slip model increase with $\phi_1$.
Furthermore, the data show that the longitudinal and transverse slip lengths are
different from each other, i.e., the flow is anisotropic, even though it
should be nearly isotropic according to the asymptotic theory, Eq.~(\ref{eq:b_small}). For striped patterns, similar
deviations from the asymptotic theory were shown to result from flow perturbations
in the vicinity of jumps in the  slip length profile~\cite{asmolov:2013}.
Here, we find that the friction and anisotropy of the flow past triangular
and trapezoidal, weakly slipping surfaces is increased, compared to the
expectation from the asymptotic theory, Eq.~(\ref{eq:b_small}). This somewhat counter-intuitive
result suggests  that similar, although perhaps weaker,
viscous dissipation and singularities  of the shear stress can also appear
if singularities in the slip length profile only appear in its first
derivative. Below we will focus on this phenomenon.

\section{Flow singularities near slipping heterogeneities}
\label{sec:edge}

Motivated by the findings from the previous section, we now discuss
the flow near slipping heterogeneity in more detail.
We remind the reader that the shear stress is known to be singular
near sharp corners of rectangular hydrophilic grooves, being proportional
to $r^{-1/3}$ for longitudinal configurations, and to $r^{-0.455}$ for
transverse configurations~\cite{wang2003} ($r$ is the distance from the
corner). In the case of a striped superhydrophobic surface, the shear
stress behaves as $r^{-1/2}$ \cite{sbragaglia.m:2007,asmolov:2012,asmolov:2013},
i.e., the singularity is stronger and creates a source of additional
viscous dissipation that reduces effective slip~\cite{asmolov:2013}.
We are now in a position to prove that similar singularities also
arise at the border between no-slip and linear local slip regions.

\begin{figure}[htbp]
  \centering
  \includegraphics[width=1.0\columnwidth]{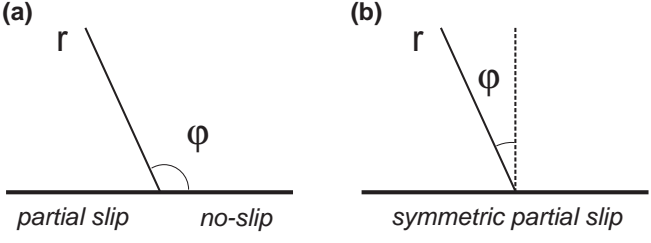}
  \caption{Polar coordinate systems for (a) trapezoidal and (b) triangular
  profiles of the local slip length. For a mathematical convenience the two coordinates differ
  in the choice of the origin of the angular variable $\varphi$.}
  \label{fig:expl}
\end{figure}

As above, we develop our theory for a longitudinal configuration only.
The solution of Laplace equation, Eq.~(\ref{eq:laplace}), can be
constructed in {the} vicinity of the edge of slipping regions
by using polar coordinates $(r,\varphi)$ \cite{wang2003,asmolov:2013}.
It is convenient to chose different coordinates for trapezoidal and
triangular textures, and to consider these two cases separately.
For a trapezoidal slip profile, we use the same coordinates
as \cite{wang2003,asmolov:2013} (illustrated in Fig.~\ref{fig:expl}(a)).
In the case of triangular textures, for symmetry reasons it is more
convenient to use the coordinates shown in Fig.~\ref{fig:expl}(b).

We begin by studying the trapezoidal slip profile.
For the trapezoid, the origin of coordinates is at
$(y_e,z_e) = ( \phi _{1}/2,0 )$, with $y=\phi _{1}/2+r\cos \varphi$ and
$z=r\sin \varphi$, $0\leq \varphi \leq \pi$, so that the n
o-slip and the slip regions correspond to $\varphi =0\ $and $\varphi=\pi$.
The slip condition (\ref{eq:bc_z0}) becomes
\begin{equation}
r\ll 1,\ \varphi =\pi :\quad u-\xi r\partial _{z}u=\xi r,  \label{bc_sl}
\end{equation}
where $\xi$ is defined as
\begin{equation}
\xi =\frac{2b_{2}}{L\left( 1-\phi _{1}-\phi _{2}\right) }, \label{definition1}
\end{equation}
and can vary from $0$ to $\infty$, depending on the local slip, $b_2/L$,
and on the value of $\phi _{1}+\phi _{2}$.
In the case of stripes, one has $\phi _{1}+\phi _{2}=1$,
resulting in $\xi \to \infty$ at finite $b_2/L$.

The solution of Eq.~(\ref{eq:laplace}) that satisfies the no-slip boundary
condition at $\varphi =0$ is given by
\begin{equation}
u=cr^{\lambda^{\parallel} }\sin \left( \lambda^{\parallel} \varphi \right) ,  \label{u_pc}
\end{equation}
and the components of the velocity gradient are
\begin{eqnarray}
  \label{du_pc}
  \partial _{z}u &=& c\lambda^{\parallel} r^{\lambda^{\parallel} -1}\cos \left[ \varphi \left( 1-\lambda^{\parallel} \right) \right] , \\
  \label{du_pc2}
  \partial _{y}u &=& -c\lambda^{\parallel} r^{\lambda^{\parallel} -1}\sin \left[ \varphi \left( 1-\lambda^{\parallel} \right) \right] .
\end{eqnarray}

The velocity at the edge is finite provided $\lambda^{\parallel} >0,$
but its gradient becomes singular when $\lambda^{\parallel} <1$.
In this case the two terms in the left-hand side of Eq. (\ref{bc_sl})
are of the same order, $r^{\lambda^{\parallel} }$, but the right-hand side
is smaller and can be neglected, since $r\ll r^{\lambda^{\parallel} }$ as $\lambda^{\parallel} <1$.
As a result Eq.~(\ref{bc_sl}) can be rewritten, taking into account (\ref{u_pc}-\ref{du_pc2}) as
\begin{equation*}
\sin \left( \lambda^{\parallel} \pi \right) -\xi \lambda^{\parallel} \cos \left[ \left( 1-\lambda^{\parallel}
\right) \pi \right] =0,
\end{equation*}
or equivalently as,
\begin{equation}
(\lambda^{\parallel}) ^{-1}\tan \left( \lambda^{\parallel} \pi \right) =-\xi .  \label{la_tr}
\end{equation}

\begin{figure}[htbp]
  \centering
  \includegraphics[width=1.0\columnwidth]{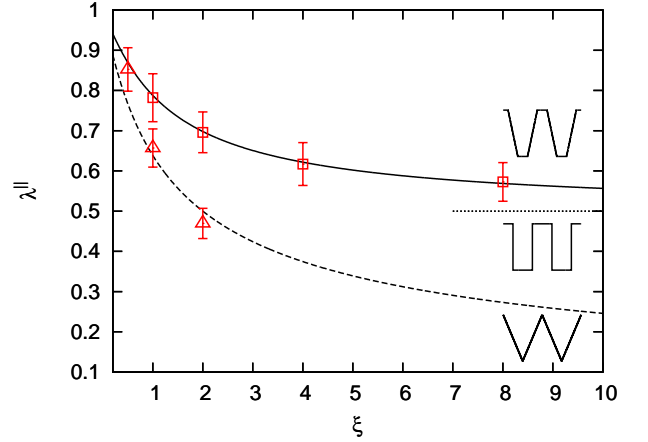}
  \caption{Exponents $\lambda^{\parallel}=1/2$ (see text for definition)
  for the trapezoid [Eq.~(\ref{u_pc}), solid line] and triangle slip profiles [Eq.~ (\ref{ur}), dashed line]
  As a reference, the exponent $\lambda^{\parallel}=1/2$ for striped surfaces is also shown
  as dotted line. Symbols are simulation data.}
  \label{fig:lam}
\end{figure}

The theoretical prediction of Eq.~(\ref{la_tr}) for the behavior of $\lambda^{\parallel}$
as a function of $ \xi $ is shown in Fig.~\ref{fig:lam} (solid curve).
Note that $\lambda^{\parallel} $ decays monotonically from
$\lambda^{\parallel} ( \xi \rightarrow 0 ) \rightarrow 1$ down to
$\lambda^{\parallel} ( \xi \rightarrow \infty ) \rightarrow 1/2$.
In other words, for any finite $\xi $, we have $\lambda^{\parallel} <1,$
and the shear stress becomes singular.
As shown in Fig. \ref{fig:lam} these theoretical predictions are in excellent
quantitative agreement with the simulation results, where we have measured the slip
velocity along the surface and then obtained $\lambda^{\parallel}$ by using Eq.~(\ref{u_pc}).
Altogether the theoretical and numerical results confirm that the singular is weaker
for trapezoidal profiles than for stripes, such that the viscous dissipation is smaller.
This explains qualitatively the results presented in Fig.~\ref{fig:phi}, where the
slip length deviates more strongly from the area-averaged slip length on striped
than on trapezoidal textures.

In the theoretical analysis of the triangular slip profile ($\phi_1=0$) it is convenient to shift the origin of the angular variable
$\phi$ compared to the trapezoidal case (see Fig.~\ref{fig:expl}), so that $z=r\cos
\varphi ,$ $y=r\sin \varphi ,$ $\left\vert \varphi \right\vert \leq
\pi /2.$

The even solution of the Laplace equation (\ref{eq:laplace}) is
\begin{equation}
u=cr^{\lambda^{\parallel} }\cos \left( \lambda^{\parallel} \varphi \right),  \label{ur}
\end{equation}
and the components of the velocity gradient are then%
\begin{eqnarray}
\partial _{z}u &=& c\lambda^{\parallel} r^{\lambda^{\parallel} -1}\cos \left[ \varphi \left( 1+\lambda^{\parallel} \right) \right] ,   \label{dzu} \\
\partial _{y}u &=& -c\lambda^{\parallel} r^{\lambda^{\parallel} -1}\sin %
\left[ \varphi \left( 1+\lambda^{\parallel} \right) \right] .
\end{eqnarray}

The slip condition (\ref{eq:bc_z0}) at $\varphi =\pm \pi /2$ can be rewritten as%
\begin{equation}
cr^{\lambda^{\parallel} }\left\{ \cos \left( \lambda^{\parallel} \frac{\pi }{2}\right) - \xi
\lambda^{\parallel} \cos \left[ \left( 1+\lambda^{\parallel} \right) \frac{\pi }{2}\right] \right\}
=\xi r,
\end{equation}
with
\begin{equation}
\xi =2b_{2}/L.\label{definition2}
\end{equation}

Applying the condition $\lambda^{\parallel} <1$, we can neglect the right-hand side to obtain
\begin{equation}
\lambda^{\parallel} \tan \left( \lambda^{\parallel} \frac{\pi }{2}\right) =\xi ^{-1}.
\label{la_z}
\end{equation}
The numerical solution of this equation is included in  Fig.~\ref{fig:lam} (dashed curve),
and supported by simulation data. We see that $\lambda^{\parallel} $ again decays
monotonically from $\lambda^{\parallel} ( \xi \rightarrow 0 ) =1$, but at large $\xi$ the solution of Eq.~(\ref{la_z}) is
\begin{equation*}
  \lambda^{\parallel}=\left( \frac{2 }{ \pi \xi }\right) ^{1/2}\ll 1,
\end{equation*}
i.e., different from expected for a trapezoidal profile, where $\lambda^{\parallel}$ would tend to 0.5.
This implies that for given $\xi$, the triangular slip profile always gives
a stronger singularity compared to a trapezoid slip profile.
At small $\xi$, the singularity is weaker than for a striped surface.

A striking conclusion from our analysis is that non-smooth surface textures
with continuous local slip can generate stronger singularities in flow and shear stress
than surfaces with a discontinuous, piece-wise constant slip length profile.
Indeed, we have $\lambda^{\parallel} > 1/2$ for $\xi>2$ (or $b_2>L$), i.e.,
the singularity for a triangular texture becomes stronger than for stripes
and should significantly reduce the longitudinal effective slip.
This is likely a reason why the effective slip in Fig.~\ref{fig:slip}(b) grows only
weakly at very large $b_2/L$. Coming back to Fig.~\ref{fig:phi}, we remark that
the results for triangular profile there correspond to $\xi=0.5$ and
$\lambda^{\parallel} \simeq 0.77$, i.e., the singularity is weaker in this
case than for stripes with $\lambda^{\parallel} =0.5$.
The singularities for trapezoidal profiles may be even weaker:
The exponents vary from $\lambda^{\parallel} \simeq 0.87$ to
$\lambda^{\parallel} \simeq 0.5$. Note however that since the no-slip area remains
the key factor determining the effective slip,
$b_{\rm eff}^{\parallel}$ always decreases with $\phi _{1}$.


To calculate the exponents $\lambda^{\perp}$ for a transverse configuration,
one can use the relation between the velocity fields in eigendirections~\cite{asmolov:2012}:%
\begin{gather}
v=\frac{1}{2}\left( u_{d}+z\frac{\partial u_{d}}{\partial z}\right) ,\quad
w=-\frac{z}{2}\frac{\partial u_{d}}{\partial y},  \label{vw} \\
p=-\frac{\partial u_{d}}{\partial y},  \label{p}
\end{gather}
where $u_{d}\left( y,z\right) =u\left[ y,z,2b\left( y\right) /L\right] $ is
the velocity field for the longitudinal pattern with a twice larger local slip
length.
It follows from Eqs.~(\ref{vw}) and (\ref{p}), that the velocity
gradients and the pressure in the transverse configuration for both trapezoidal and triangular textures have
stronger singularities at the edges of slipping regions than those for the
longitudinal configuration:
\begin{equation*}
\frac{\partial v}{\partial z},\ \frac{\partial w}{\partial z},\ p\sim
r^{\lambda^{\perp}-1},
\end{equation*}
where $\lambda^{\perp}=\lambda^{\parallel} \left( 2\xi \right) <\lambda^{\parallel} \left( \xi \right) $.
This leads to larger viscous dissipation. As a result the anisotropy of the slip-length tensor
is finite even at small $b_2/L$ and is approximately the same for triangular, trapezoidal and stripe profiles (
see Fig. \ref{fig:phi}). For example, the eigenvalues shown there for a triangular texture
correspond to $\lambda^{\parallel} \simeq 0.77$ and $\lambda^{\perp} \simeq 0.64$. Therefore, in the transverse configuration even for a shallow triangular texture, $b_{2}>L/2$, the singularity becomes stronger than for stripes ($\lambda ^{\perp }>1/2)$.

\section{Optimization of the effective slip}

Based on the results obtained in the previous sections, we now discuss how to
optimize the effective slippage, and what maximum slip lengths may actually
be expected, taking into account the actual limitation in surface engineering.

\begin{figure}[tbp]
  \centering
  \includegraphics[width=1.0\columnwidth]{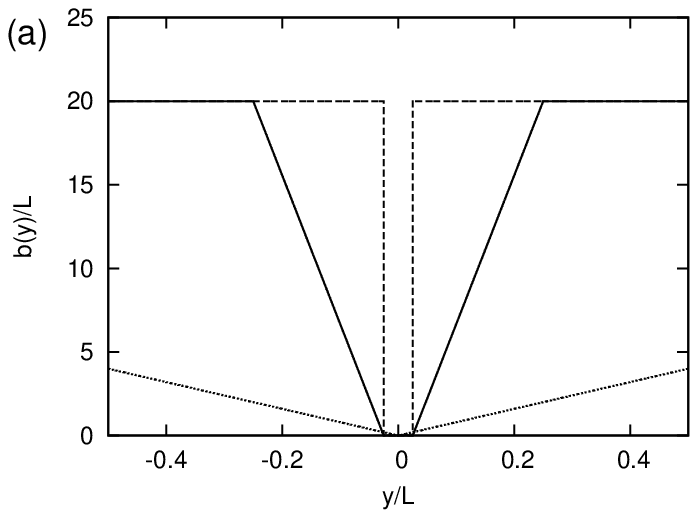}\\
  \includegraphics[width=1.0\columnwidth]{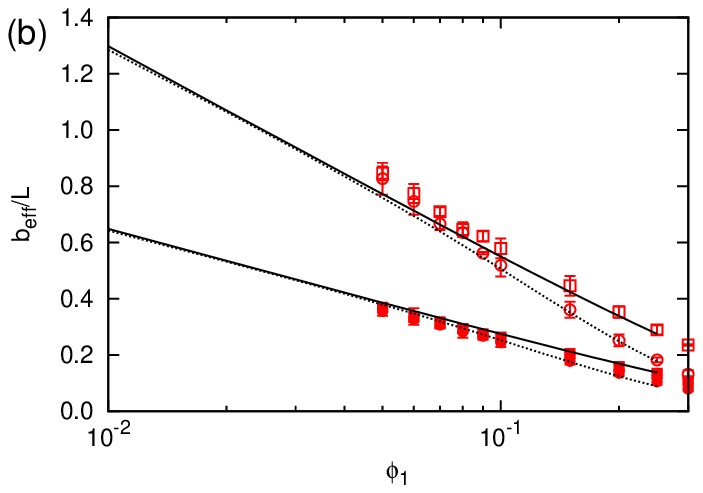}
  \caption{ (a) Profiles of local slip length resulting in the same $b_{\rm eff}^{\parallel}$,
(b) Effective slip lengths for stripe (solid lines and square symbols) and trapezoidal
 (dotted lines and circular symbols) local slip profiles. The top and bottom lines correspond to
$b_{\rm eff}^{\parallel}$ and $b_{\rm eff}^{\perp}$, respectively.}
\label{fig:trap_nar}
\end{figure}

Very large effective slip lengths were recently found for sinusoidal
profiles $b(y)$ with a no-slip point \cite{asmolov:2013b}. If we request
that the local slip pattern includes a no-slip area, we also find here
that the largest values of $b_{\rm eff}$ are obtained for
triangular profiles, where the no-slip area reduces to just one point.
However, real textures always have finite contact area with liquid.
The effective slip length of a trapezoidal texture with large $b_2/L$
is close to that of stripes [see Fig.~\ref{fig:slip}(a)].
In this final section, we therefore focus on the case of a trapezoidal
profile with small $\phi_1$, where both $b_2/L$ and the slope of the inclined
regions is large. Such a texture should lead to large $b_{\mathrm{eff}}$.
To find one-dimensional textures which optimize the effective forward
slip we consider profiles of the following form
\begin{equation}
  \label{eq:trap2}
  \frac{b(y)}{L}=
  \begin{cases}
    0, & |t|\leq \frac{\phi_1 }{2} \\
    \frac{2|t|-\phi_1 }{9\phi_1 ^{2}}, & \frac{\phi_1 }{2}<|t|\leq 5\phi_1 \\
    \phi_1 ^{-1}, & 5\phi_1 <|t|\leq \frac{1}{2}, \ \phi_1 <0.1%
  \end{cases}%
\end{equation}
The bottom base of the trapezoid for such textures is wide, $10\phi_1$, [see Fig.~\ref{fig:trap_nar}(a)].
For comparison, we also consider stripes with the same maximal local slip length,
$b_2/L=\phi_1 ^{-1}$:
\begin{equation}
\frac{b(y)}{L}=%
\begin{cases}
0, & |t|\leq \frac{\phi_1 }{2} \\
\phi_1 ^{-1}, & \frac{\phi_1 }{2}<|t|\leq \frac{1}{2}%
\end{cases}%
\end{equation}
Both profiles are characterized by only one parameter $\phi_1$.

The simulation and numerical results for $b_{\rm eff}^{\parallel}$ and
$b_{\rm eff}^{\perp}$ are shown in Fig.~\ref{fig:trap_nar}(b).
The effective slip lengths for the trapezoid and stripe profiles are nearly equal for $\phi_1 <0.1$.
For example, $b_{\rm eff}^{\parallel}$ for the stripe profile with $\phi_1 =0.05$
is only slightly larger than the corresponding value for the trapezoid profile
with the same $\phi_1$ and wide bottom base [see Fig.~\ref{fig:trap_nar}(a)].
Note that such narrow trapezoids with wide bottom base allowed researchers to achieve
a very large effective slip in recent experiments~\cite{choi.ch:2006}. The same
value of $b_{\rm eff}^{\parallel}$ can be achieved with a triangular profile
with $b_{\mathrm{max}}/L=4$. Thus the effect of the inclined regions is relatively
small for large slopes, and the main contribution to the effective slip length
stems from the no-slip region and the singularity of the shear stress.

\section{Concluding Remarks}

In conclusion, we have presented simulation and numerical results and some asymptotic law
analysis that allowed us to assess the frictional properties of superhydrophobic surfaces
as a function of the texture geometric parameters.
We have focused on one-dimensional superhydrophobic surfaces with trapezoidal grooves
and systematically investigated the effect of several parameters, such as the
maximum local slip and the heterogeneity factor.
The slip properties on a trapezoidal surface were compared with two limiting cases:
striped and triangular slip length profiles. Our simulation and numerical results
suggested that the effective slip length is mainly determined by
(i) the area fraction of the no-slip region and
(ii) the singularity of the shear stress at the no-slip edge.
The first factor is obvious; a smaller fraction of no-slip area results in less
friction of the surface, which in turn leads to large slip length.
The second effect is less intuitive, especially for a continuous slip profile.
Here we demonstrated that the singularity develops for the trapezoidal and triangular
local slip profiles, and in the latter case can be even stronger than for
discontinuous slip of striped textures.
The singularity is stronger for a transverse configuration in comparison to a longitudinal
configuration, and as a result, it always contributes to the flow anisotropy.
Our results can guide the design of superhydrophobic surfaces
for micro/nanofluidics.

\begin{acknowledgments}
This research was supported by the RAS through its priority program
``Assembly and Investigation of Macromolecular Structures of New
Generations'', and by the DFG through SFB TR6 and SFB 985.
The simulations were carried out using computational resources at the
John von Neumann Institute for Computing (NIC J\"ulich), the High
Performance Computing Center Stuttgart (HLRS) and Mainz University (MOGON).
\end{acknowledgments}




\end{document}